\documentclass[aps,prl,twocolumn,notitlepage,superscriptaddress,showpacs]{revtex4-1}

\usepackage{amsmath,latexsym,amssymb,bm}
\usepackage{hyperref}
\usepackage{color}
\usepackage{graphicx,subfigure}
\usepackage{multirow}
\usepackage{color}
\begin{document}

\title{Decoy-State Quantum Key Distribution over a Long-Distance High-Loss\\ Underwater Free-Space Channel}

\author{Cheng-Qiu Hu}
\affiliation{Center for Integrated Quantum Information Technologies (IQIT), School of Physics and Astronomy and State Key Laboratory of Advanced Optical Communication Systems and Networks, Shanghai Jiao Tong University, Shanghai 200240, China.}
\affiliation{CAS Center for Excellence and Synergetic Innovation Center in Quantum Information and Quantum Physics, University of Science and Technology of China, Hefei, Anhui 230026, China}

\author{Zeng-Quan Yan}
\affiliation{Center for Integrated Quantum Information Technologies (IQIT), School of Physics and Astronomy and State Key Laboratory of Advanced Optical Communication Systems and Networks, Shanghai Jiao Tong University, Shanghai 200240, China.}
\affiliation{CAS Center for Excellence and Synergetic Innovation Center in Quantum Information and Quantum Physics, University of Science and Technology of China, Hefei, Anhui 230026, China}

\author{Jun Gao}
\affiliation{Center for Integrated Quantum Information Technologies (IQIT), School of Physics and Astronomy and State Key Laboratory of Advanced Optical Communication Systems and Networks, Shanghai Jiao Tong University, Shanghai 200240, China.}
\affiliation{CAS Center for Excellence and Synergetic Innovation Center in Quantum Information and Quantum Physics, University of Science and Technology of China, Hefei, Anhui 230026, China}

\author{Zhan-Ming Li}
\affiliation{Center for Integrated Quantum Information Technologies (IQIT), School of Physics and Astronomy and State Key Laboratory of Advanced Optical Communication Systems and Networks, Shanghai Jiao Tong University, Shanghai 200240, China.}
\affiliation{CAS Center for Excellence and Synergetic Innovation Center in Quantum Information and Quantum Physics, University of Science and Technology of China, Hefei, Anhui 230026, China}

\author{Heng Zhou}
\affiliation{Center for Integrated Quantum Information Technologies (IQIT), School of Physics and Astronomy and State Key Laboratory of Advanced Optical Communication Systems and Networks, Shanghai Jiao Tong University, Shanghai 200240, China.}
\affiliation{CAS Center for Excellence and Synergetic Innovation Center in Quantum Information and Quantum Physics, University of Science and Technology of China, Hefei, Anhui 230026, China}

\author{Jian-Peng Dou}
\affiliation{Center for Integrated Quantum Information Technologies (IQIT), School of Physics and Astronomy and State Key Laboratory of Advanced Optical Communication Systems and Networks, Shanghai Jiao Tong University, Shanghai 200240, China.}
\affiliation{CAS Center for Excellence and Synergetic Innovation Center in Quantum Information and Quantum Physics, University of Science and Technology of China, Hefei, Anhui 230026, China}

\author{Xian-Min Jin}
\affiliation{Center for Integrated Quantum Information Technologies (IQIT), School of Physics and Astronomy and State Key Laboratory of Advanced Optical Communication Systems and Networks, Shanghai Jiao Tong University, Shanghai 200240, China.}
\affiliation{CAS Center for Excellence and Synergetic Innovation Center in Quantum Information and Quantum Physics, University of Science and Technology of China, Hefei, Anhui 230026, China}
\address{xianmin.jin@sjtu.edu.cn}


\maketitle
\textbf{Atmospheric free space and fiber have been widely exploited as the channels for quantum communication, and have enabled inter-continent and inter-city applications. Air-sea free-space channel, being capable of linking the satellite-based quantum resource and underwater vehicle, has now become the last piece of the puzzle in building global quantum communication network. However, long-distance quantum communication penetrating water up to tens to hundreds of meters is extremely challenging due to the inevitable high loss. Here, we present an experimental demonstration of underwater decoy-state quantum key distribution against high loss, meanwhile keep a low quantum bit error rate less than 2.5\% for different distances. By directly modulating blue-green lasers at a high speed of 50MHz and decoy-state protocol, we are able to for the first time reach a long-distance quantum key distribution that is unconditionally secure and can enable real-life air-sea quantum communication tasks. The demonstrated distance, even in coastal water of Jerlov types 2C, is up to 30 meters, about one-order improvement over the proof-in-principle demonstrations in previous experiments, and the channel loss is equivalent to 345-meter-long clean seawater of Jerlov type I, representing a key step forward to practical air-sea quantum communication.}

\section{Introduction}
Quantum key distribution (QKD), as an ingenious combination of traditional cryptography and quantum mechanics, allows remote individuals to share secrets with unconditional security. The first QKD protocol (known as BB84) was proposed in mid 1980s and implemented over a 32-cm-long free-space air channel \cite{bennett1984proceedings}, after which worldwide researches ensued\cite{lo1999unconditional,shor2000simple,gottesman2004security}. So far, satellite-based quantum resource \cite{ren2017ground,liao2017satellite,takenaka2017satellite} and fiber-linked quantum network \cite{kimble2008quantum,sasaki2011field,lucamarini2018overcoming} are applicable for intercontinental and intercity QKD, leaving only the links between the airborne quantum resource and the underwater vehicles unbuilt.

Compared with the 3dB attenuation by aerosphere\cite{jin2010experimental}, only tens of meters of underwater free space will cause orders of magnitude higher photon loss. The core problem of air-sea QKD is to overcome the huge channel loss and still possessing unconditional security. Fortunately, the proposal of decoy-state method\cite{wang2005beating,ma2005practical} makes it possible for practical QKD over high channel loss up to 40-50dB and at the same time remain unconditionally secure\cite{wang2013direct,nauerth2013air,fedrizzi2009high}. It has been proved that photonic polarization and entanglement can be maintained well through water channel\cite{ji2017towards,hu2019transmission}. The progresses have also been made in theoretical analysis\cite{gariano2019theoretical,raouf2020performance,zhao2019performance} and underwater experiments using polarization encoding \cite{ji2017towards,hu2019transmission,zhao2019experimental,li2019proof}and twisted photons\cite{chen2020underwater,bouchard2018quantum,hufnagel2019characterization}, however, still are limited as proof-of-principle demonstration, in which the longest water channel is merely several meters. 
\begin{figure*}[!]
\includegraphics[width=2\columnwidth]{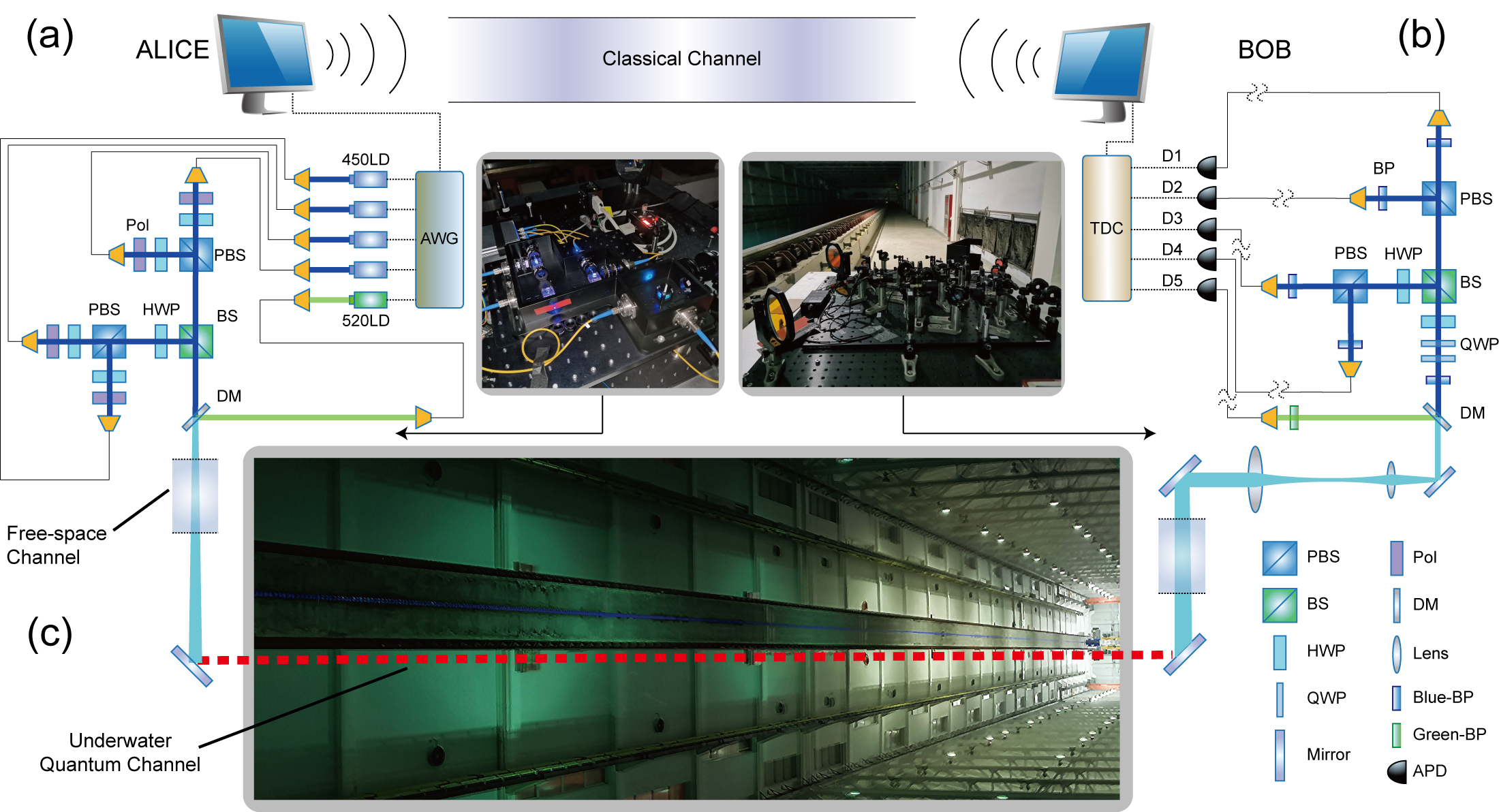}
\renewcommand{\figurename}{Fig.}
\caption{\textbf{Experimental setup. } \textbf{(a) (b),} Illustration and photograph of the Alice and Bob end. At the Alice end, four blue laser diodes (450 LDs) are driven by the arbitrary wave generator (AWG) to emit short pulses (3ns, 50MHz). Their light beams are overlapped at the (polarization) beamsplitters (BS, PBS). A green laser diode (520 LD) is used for time synchronization. A dichroic mirror (DM) is used to combine the 450nm and the 520nm beams. The merged laser beam goes through a 2-m-long free-space channel and is guided to the underwater channel by two mirrors. At the Bob end, the laser beam is collimated and led to the BB84 decoding module by mirrors and lens. Blue and green bandpass filters (BP) are used for spectrum filtering. Five single photon detectors (D1-D5) transform the photon pulses into electrical signals for the time-digital convertor (TDC). Alice and Bob use the local area network for classical communication. \textbf{(c),} The photograph of our experimental site. The underwater channel is denoted by the dashed red line. HWP: half-wave plate, QWP: quarter-wave plate, Pol: polarizer, APD: avalanche photodiode.}
\label{fig1}
\end{figure*}

In this work, we successfully implement quantum key distribution through a long and high-loss air-water-air channel with an average quantum bit error rate (QBER) less than 2.5\%. The advantages of our self-developed blue-green QKD transmitter at the wavelength of 450nm with high modulation speed and 3-intensity decoy-state protocol make it possible for our system to tolerate high total loss up to 35dB. The distance of underwater part for the first time reaches 30 meters and the channel loss is 27dB, as high as in 345-m-long clean seawater of Jerlov type I. The underwater distance is far above the forbidden region of traditional radio frequency signal,  which is very close in building practical air-sea quantum links.

\begin{figure*}[!]
\includegraphics[width=2\columnwidth]{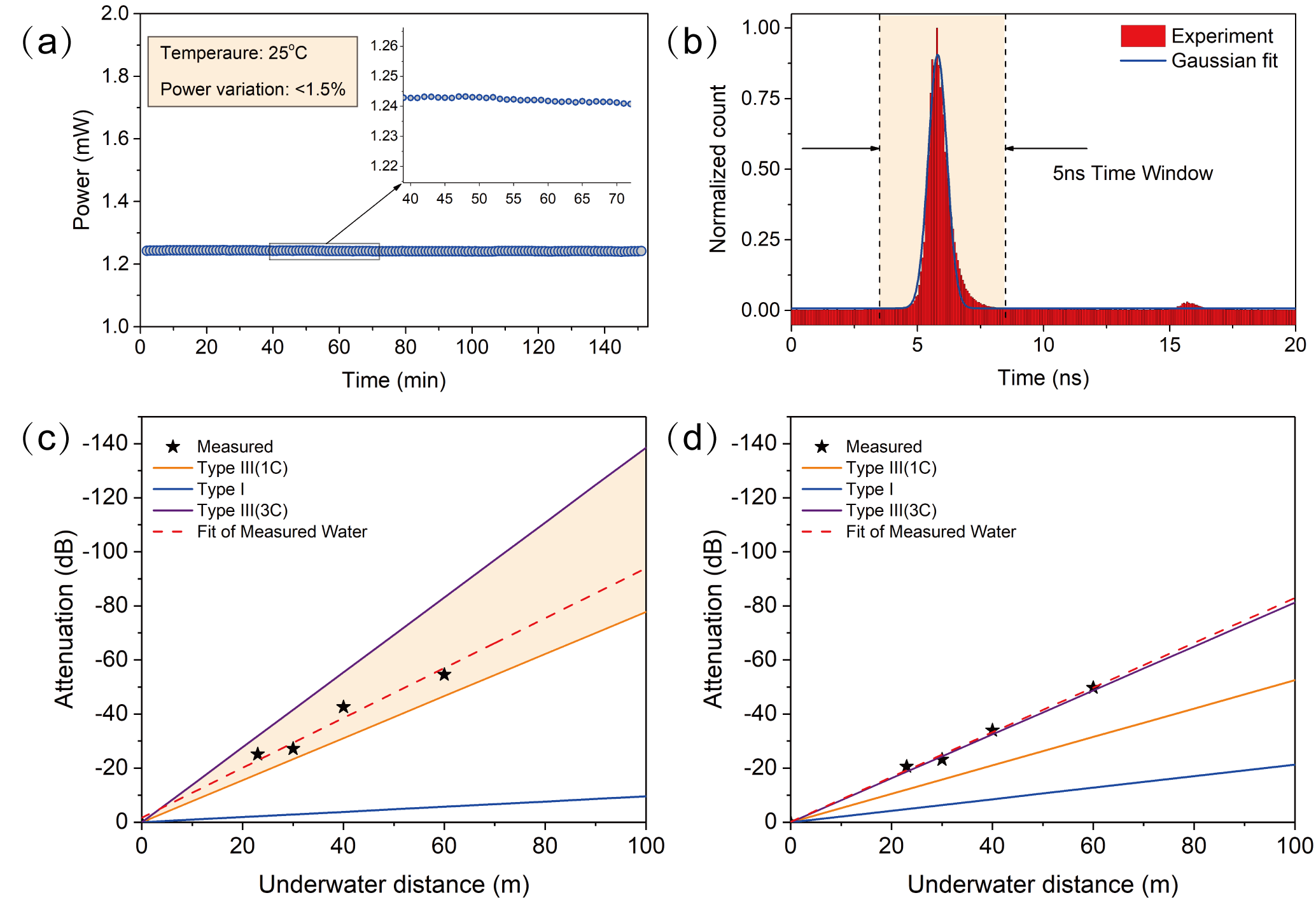}
\renewcommand{\figurename}{Fig.}
\caption{\textbf{(a), The power stability tested over 150 minutes.} The average power is 1.242mW and the variation is less than 0.017mW (1.5\%). \textbf{(b), Temporal shape of the the signal pulse.} We use the single photon detectors and time-digital convertor to record the photon counts and  reconstruct the signal pulse. The pulse width is $\sim$3ns and the time window we pick in the experiment is 5ns. \textbf{(c)(d), Optical attenuation in different types of water.} The measured photon loss (denoted by the asterisks) in the experiment is between Jerlov type III(1C) and Jerlov type III(3C). The corresponding optical wavelength: (c) 450nm, (d) 520nm.}
\label{fig2}
\end{figure*}

\section{Experimental setup}
In order to reach a long-distance underwater channel similar to a natural field situation, we choose the large-scale marine test platform to implement our experiment. As shown in Fig. 1(c), the whole platform is in a semi-open environment with a water pool measured 300m long, 16m wide and 10m deep, which makes it close to a real field condition. Considering the impact of air-water interface on quantum communication, our experiment is designed to be the air-water-air way that involves 2m long free-space channels in both ends and the underwater channel (denoted by the dashed red line in Fig 1(c)) between the air channels. The photon incident angle at the air-water interface is set to near 90 degree. 

For the transmitter’s end (referred as Alice), we designed a compact transmitter system for generating quantum signals. The Alice end mainly contains two parts: a self-made BB84-encoding module of size 30cm$\times$30cm (see Fig .1(a)) and an self-assembled laser system. We use four blue laser diodes to prepare decoy states and a green laser diode for time synchronization. An 8-channel arbitrary wave generator (AWG) producing key patterns is linked to the laser modulating port. 

For the receiver’s end (referred as Bob), we use a two-lens system to collimate the laser beam and detect the photons using five silicon single photon detectors (D1-D5) after the polarization measurement. As shown in Fig .1(b), the BB84-decoding module mainly consists of a 50:50 beam splitter (BS), a half-wave plate (HWP) and two polarization beam splitters (PBS).

\section{Fast and polarization encoding in blue-green window}
Like the wavelength around 800nm is usually used in atmosphere, the blue-green optical window (generally refers to 400-500nm) is preferred involving underwater channel, wherein photons experience less loss and therefore can be utilized for cross-medium communication. Here, we choose the center wavelength at 450nm and 520nm for signal and timing pulses respectively. Based on the results of our earlier works \cite{jin2010experimental,ji2017towards,hu2019transmission}, polarization encoding is suitable for QKD in underwater free space as well as in atmosphere because of its high fidelity through these channels. A big challenge is that unlike telecom wavelength, where polarization modulating (PM) technology (e.g. high-speed electro-optic modulator) is mature, there is no such effective PM modulator for the blue-green region in prior art. 

To produce the high speed and narrow pulses required by our experiment, we thus use voltage signals to directly modulate four independent laser diodes at the repetition rate up to 50MHz. Each laser then goes through a half-wave plate and a polarizer to precisely normalize its intensity and define its polarization. The laser diodes are integrated with a closed-loop temperature control so that their output powers and spectrum are stabilized. The stability performance of our self-assembled laser system and the optical signal pulse are presented in Fig. 2(a)(b). The power variation of our laser system is less than 1.5\% over the test time of two and half hours. The temporal shape of the signal pulse we obtained is near Gaussian and its width is narrow enough to be covered by the 5ns time window for the signal coincidence and noise filtering.

\section{Experiment procedure and results}
We first characterize the underwater channel loss in several different distances, i.e. 23m, 30m, 40m and 60m with 450nm and 520nm lasers. By comparing the output power with its initial one, we obtained the water attenuation listed in Table .I. Referring to Jerlov’s definition of water types\cite{jerlov1968optical,arnone2004evolution}, we then plot the results in Fig .2(c)(d), from which we can see that our water is between Jerlov type III(1C) and Jerlov type III(3C)\cite{solonenko2015inherent}. 
\begin{table}[htb!]
\includegraphics[width=1\columnwidth]{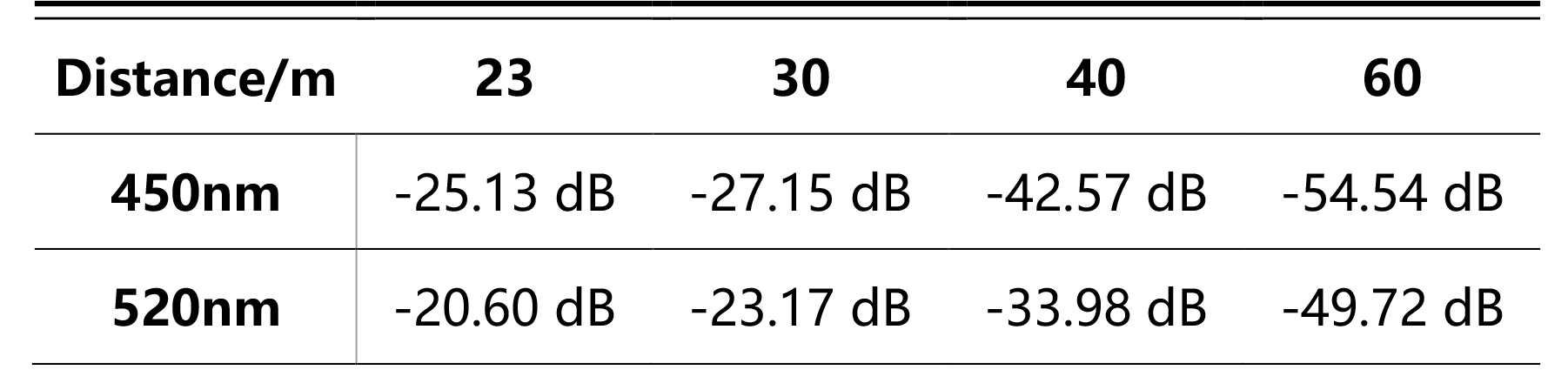}
\renewcommand{\figurename}{Table.}
\caption{\textbf{Measured attenuation of the water in the experiment.}}
\label{table1}
\end{table}
\begin{table}[!]
\includegraphics[width=1\columnwidth]{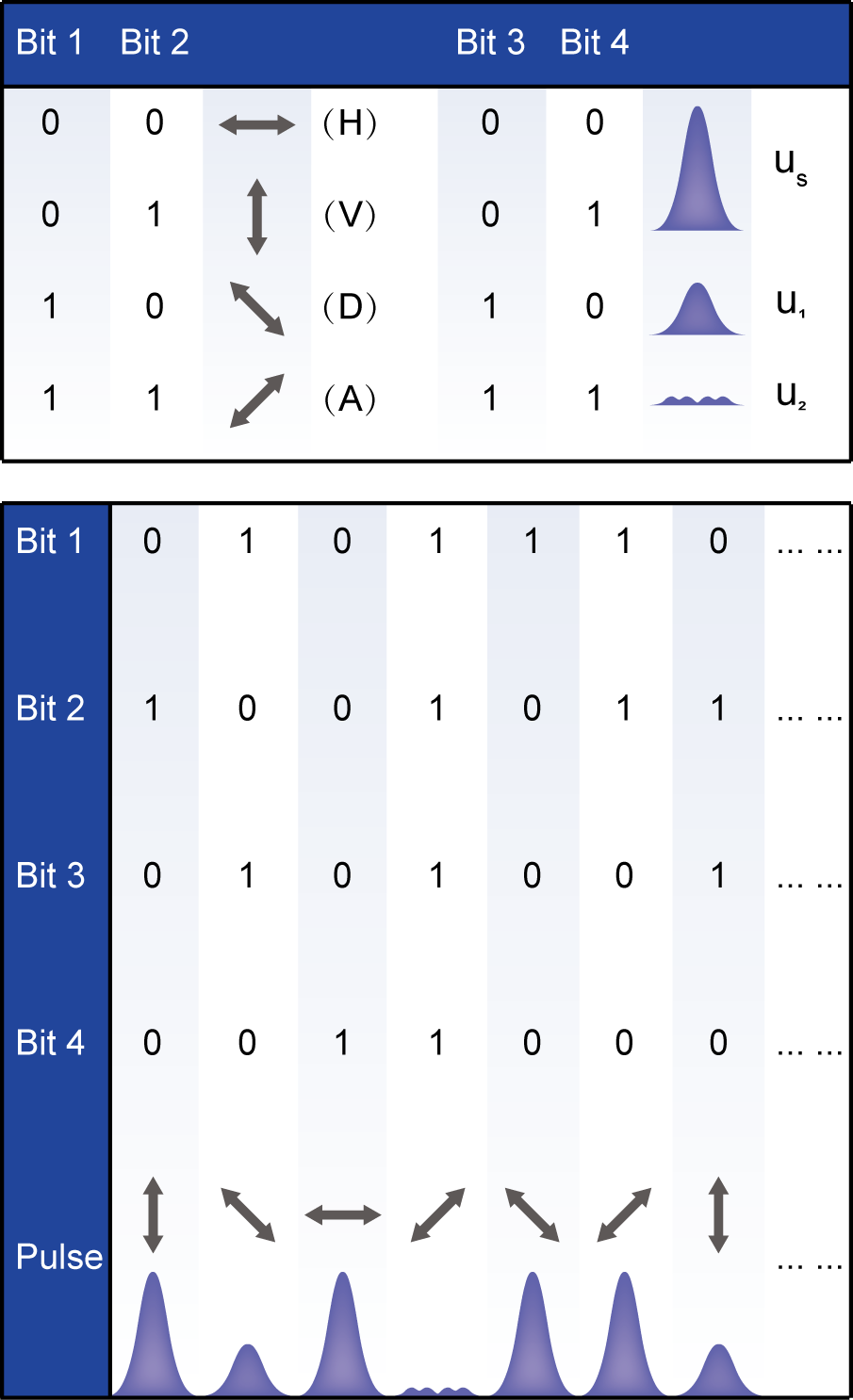}
\renewcommand{\figurename}{Fig.}
\caption{\textbf{Illustration of the encoding protocol.} The bit1 and bit2 decide the polarization states. At the same time, the bit3 and bit4 decide the pulse intensity. For example, the combination of the 4-bit random number \{0111\} lead to a high-intensity pulse $u_s$ of the state V. Some other combinations are listed in the bottom half of the table.}
\label{table2}
\end{table}
\begin{table*}[t]
\includegraphics[width=2\columnwidth]{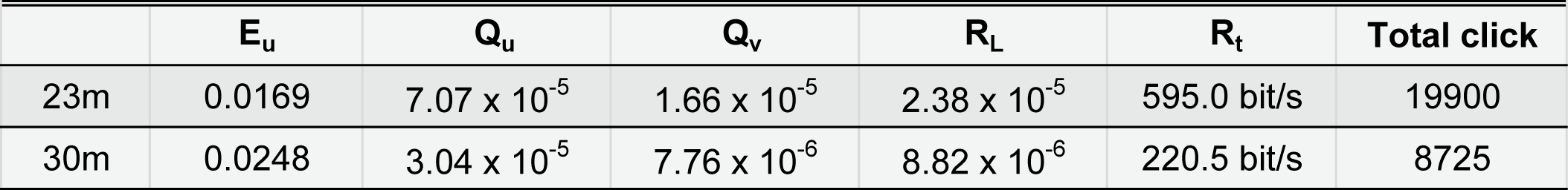}\renewcommand{\figurename}{Table.}
\caption{\textbf{Key parameters of the QKD experiment.} $E_u$ is the quantum bit error rate (QBER). $Q_u$ and $Q_v$ are the gains of the signal states $u_s$ and decoy states $u_1$. $R_L$ is the key generation rate. $R_t$ is the bit rate of the final keys.}
\label{table2}
\end{table*}

Taking the above parameters into consideration, the decoy-state BB84 protocol, which possesses the immunity to photon-number-splitting (PNS) attack and thus enables secure QKD using weak coherent laser source over high-loss channel, is a reasonable option for our experiment. By randomly mixing several decoy states of different intensity with certain proportion at the source of the transmitter, secret sharers make sure any PNS eavesdropping can be detected. Here, we utilize attenuators and adjust independent HWPs to get accurate three intensity decoy states, of which the average photon numbers per pulse are high $u_s=0.9$, moderate $u_1=0.3$ and vacuum state $u_2=0$. The mixing ratio are 50\%, 25\% and 25\% for $u_s$, $u_1$ and $u_2$ respectively. 

We use the four polarization states: horizontal (H), vertical (V), +45 (D) and -45 (A) equiprobably to encode the secret keys. For clarity, we illustrate the encoding protocol in Table .II. As shown in Table .II, for each round a pre-prepared 4-bit random number is consumed to drive the four lasers and determine which polarization and intensity level the pulse will be. A 520nm pulsed laser with a repetition rate of 500KHz serves as the beacon light and the time synchronization signal. 

At the Bob end, we use a dichroic mirror (DM) to coarsely separate the photons of two different wavelengths, after which most of the 520nm photons are uncoupled and detected by the D5. Then the 450nm photons are further purified by a blue bandpass filter (center wavelength at 445nm, band width 20nm). Apart from the BB84 decoding module, there are two QWPs and an HWP functioning as polarization compensation, which we accurately adjust to calibrate the whole system before the QKD process. We prepare the four polarization states: H, V, D and A at the Alice end and send them to Bob for testing. Fig. 3 shows the fidelities of these states after the calibration. The average fidelity is as high as 0.982, which indicates that the system is ready for proceeding QKD.
\begin{figure}[t]
\includegraphics[width=1\columnwidth]{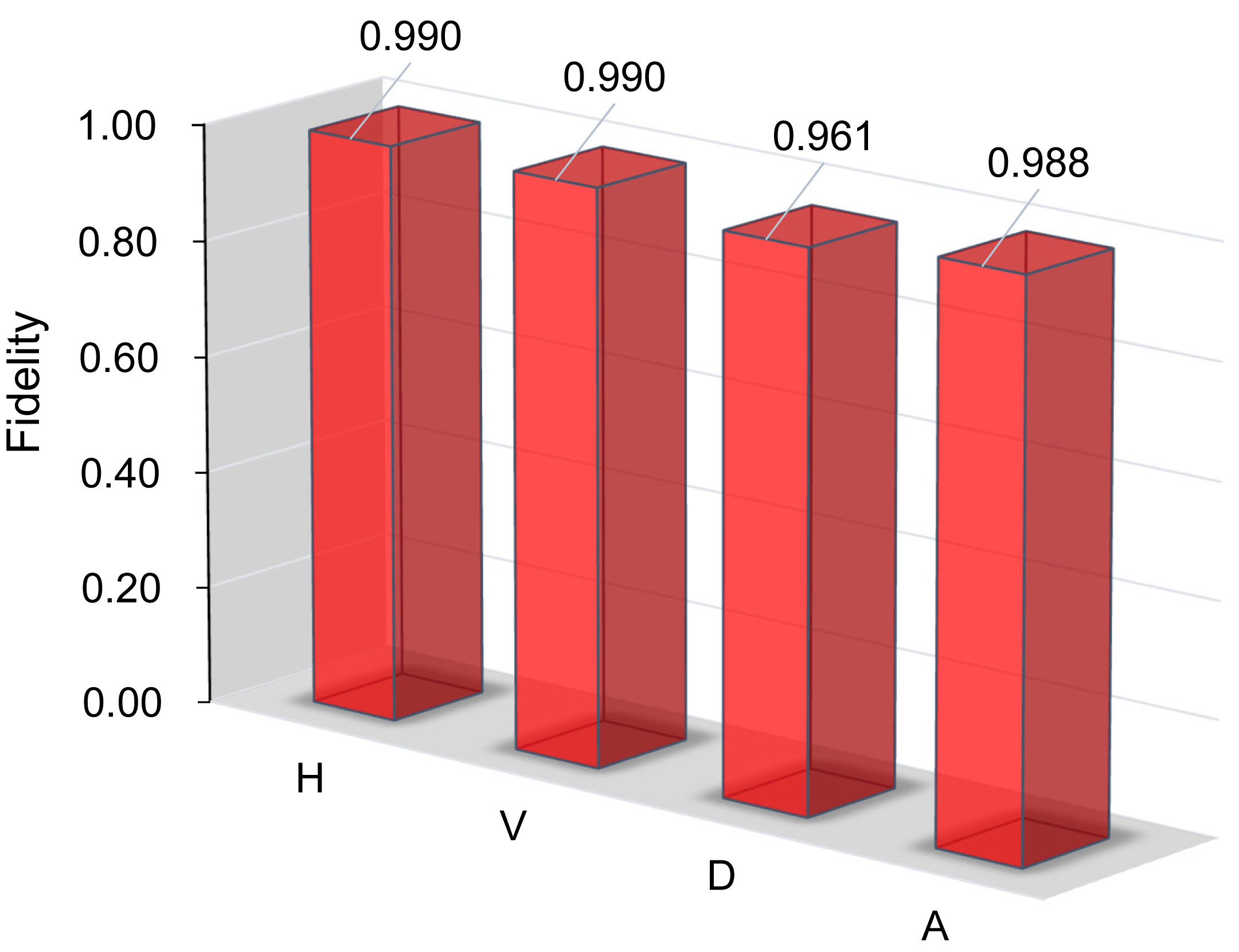}
\renewcommand{\figurename}{Fig.}
\caption{\textbf{The fidelities of the four polarization states.} High fidelities ($>$0.96) are obtained for all the four polarizaiton states after the system calibration, of which the average value is up to 0.982.}
\label{fig3}
\end{figure}
\begin{figure*}[t]
\includegraphics[width=2.1\columnwidth]{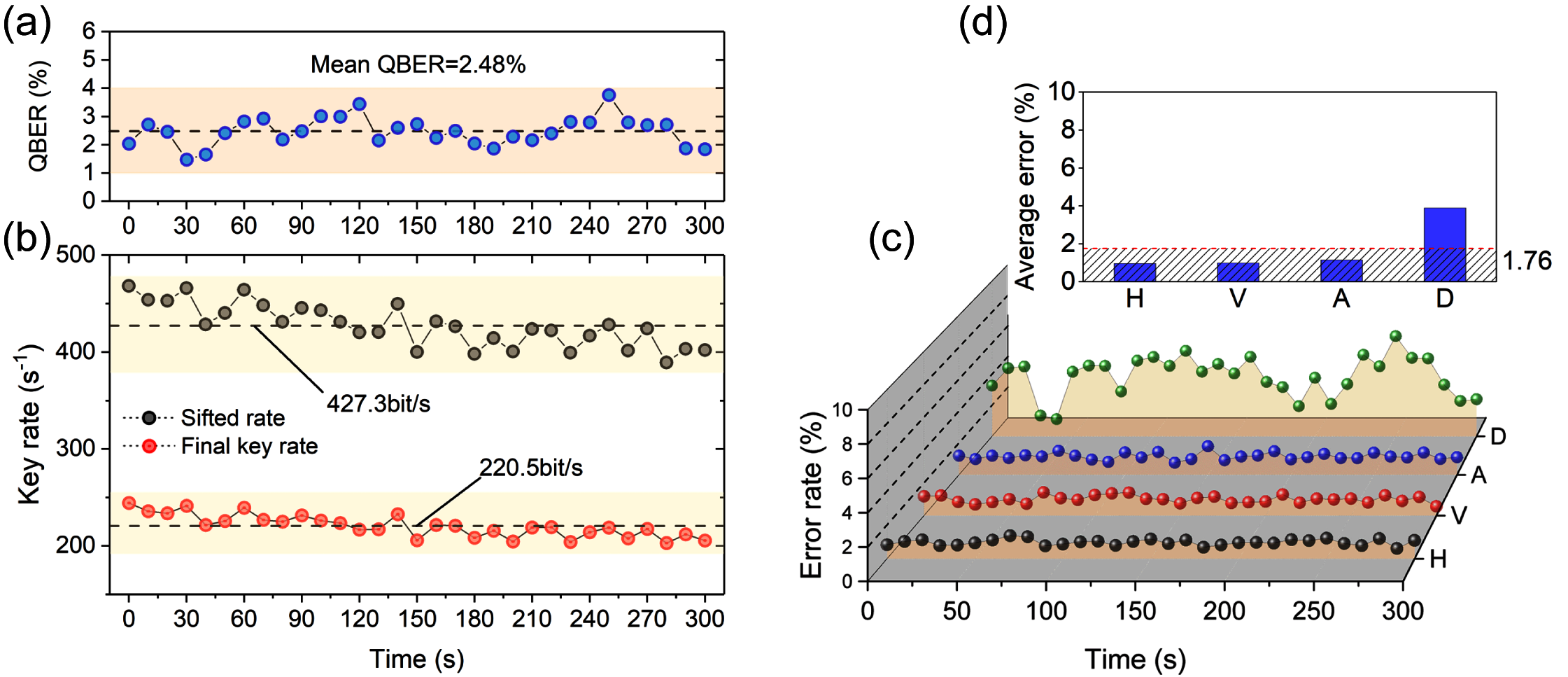}
\renewcommand{\figurename}{Fig.}
\caption{\textbf{The results of 30m underwater QKD.} \textbf{(a)(b),} The QBER and secret key rate of the first 30 rounds QKD. The obtained average QBER over the time of 300s is 2.48\%, the average sifted key (final key) rate is 427.3 bit/s (220.5 bit/s). \textbf{(c)(d),} The error rates of the four different encoding states. The error of each state over time is plotted in (c) and their average values are represented by the histograms in (d). }
\label{fig4}
\end{figure*}

The spot diameter of the light beam arriving at the Bob end after the collimation is $\sim$2mm. Through a fine optical alignment, we finally obtain a photon collecting efficiency up to 70\%. The single photon detectors we used in experiment possess an average quantum efficiency of 20\% (25\%) for 450nm (520) photons, an average dark count less than 50Hz and time jitter less than 250ps. The total loss of our system is about 35dB. By spatial and spectrum filtering, the environment background noise is suppressed to $\sim$100Hz. The photoelectric signals of D1-D5 are collected by the time-digital-convertor (TDC) and then sent to Bob’s computer. We designed a software based on MATLAB for real time post-processing, which includes base sifting, error estimation, error correction and privacy amplification process. 

 In two different distances of underwater channel (23m and 30m), we continuously run the system for 30 minutes, distributing keys for about 140 rounds. The successfully shared secret keys add up to 72.8 Kbit in 23m and 30.5 Kbit in 30m. The general experimental parameters of both distances over the first 30 rounds are shown in Table .III, from which we can see the average QBER is less than 1.7\% (2.5\%) in 23m (30m). In the 23m experiment, we obtain a maximum key rate of 715 bit/s in one single round and an average key rate $R_t$=595 bit/s over the first 30 rounds.  

We also present the detailed results of the 30m experiment in Fig .4(a)(b) to show the system performance during a consecutive period of time. The BS we use for passive base sifting possess an average sifting rate 0.489, which is very close to the ideal case 0.5. The average sifted key rate and the final key rate over the 30 rounds are 427.3 bit/s and 220.5 bit/s respectively. In addition, the individual error rates of each polarization state are analysed and plotted in Fig .4(c)(d). An average error rate of 1.76\% is obtained for the four encoding states.

The main contribution (about 70\%) of the quantum bit error rate comes from the device imperfections and imperfect polarization compensation. Other error leading factors include the dark count of single photon detectors, the environment background noise and the laser source background. 
\begin{figure*}[!]
\includegraphics[width=1.45\columnwidth]{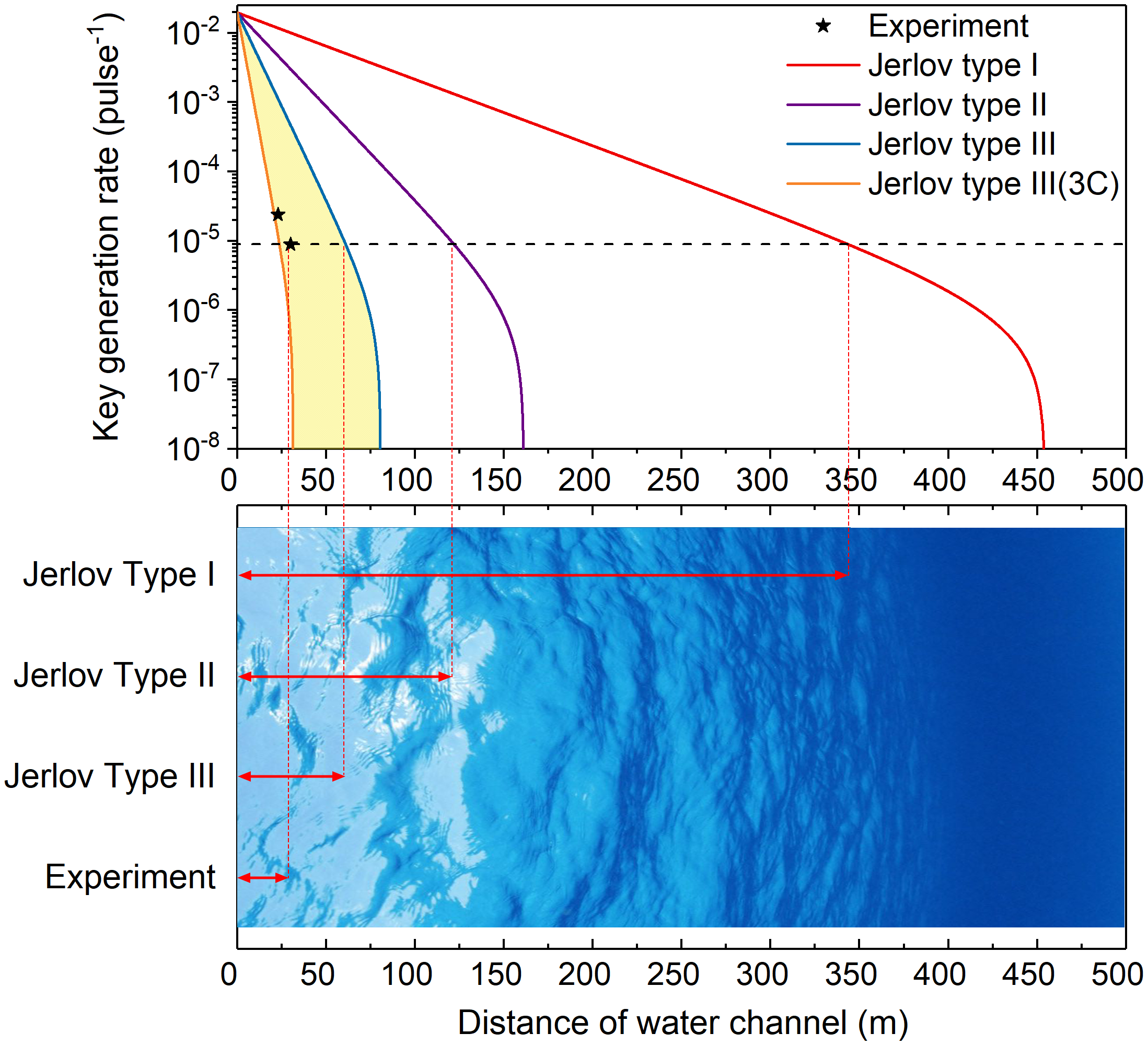}
\renewcommand{\figurename}{Fig.}
\caption{\textbf{Key generation rates in different distances of underwater channel.} The relation curves are given by the GLLP analysis combined with the decoy method and the asterisks denote our experimental results. The bottom half of the figture shows that longer distances can be achieved in different types of water, wherein photons will experience the same channel loss as in our experiment.}
\label{fig5}
\end{figure*}
\section{conclusion and discussion}
We successfully demonstrate QKD through a 30-m-long underwater channel with high loss. Our self-developed blue-green QKD system can tolerate up to 27dB water attenuation and total system loss of 35dB. As illustrated in Fig .5, the same channel loss allows quantum communication over 345-m-long (120-m-long) water channel of Jerlov type I (II) according to the GLLP\cite{gottesman2004security} analysis combined with the decoy method:\cite{ma2005practical}
\begin{equation}
R\geq q\{-Q_{u}f(E_{u})H_2(E_u)+Q_1[1-H_2(e_1)]\}
\end{equation}
where the $q$ depends on the sifting rate (in our experiment is 0.489). The $Q_1$ and $e_1$ are the gain and error rate of single-photon states, of which the lower bound and upper bound can be estimated by the decoy method. The $f(x)$ is the bidirectional error correction efficiency and the $H_2(x)$ is the binary Shannon information function. The long distance up to hundreds of meters in seawater is promising for many practical applications such as quantum links between submersibles and satellites. 

Due to the skin effect, traditional radio frequency signal (known as RF signal) can only penetrate few meters of seawater, which makes it impractical for cross-medium communication. Taking this into account, our system is also designed compatible with underwater wireless optical communication device, which makes it possible for air-sea secure communication using only optical system. 

\section{Methods}
\noindent\textbf{Optical alignment:}
One big challenge in our experiment is to realize a good optical alignment through the underwater channel. As shown in Extended Fig.1, to keep the polarization components (H, V, D, A) balance at the receiver end, perfectly overlapped spatial modes of the four signal paths must be guaranteed. Thus, we recollect the signal beams into a single-mode fiber after the BS in the transmitter. The coupling efficiency of each components can be adjusted independently to reach the same value (50\%). Firstly, we use the DM to combine the signal and the beacon beam. We tune the collimators of the beacon light to change the pointing angle and focal length so that the spatial mode overlaps the signal light over a long distance of free space (200m). Then we switch the 520LD to high-power mode and guide the combined light into water. After attenuation of the underwater channel, only the green beacon light is visible. We coarsely adjust the large-size mirrors at the transmitter end to control the pointing direction and finely tune the large-size mirrors at the receiver end so that the light is well collected by the lens system, after that the optical alignment is finished.\\

\noindent\textbf{Time synchronization:}
The real-time post processing of QKD requires Alice (the transmitter) and Bob (the receiver) to share a common time frame, which means the bit sequence detected by Bob corresponding to the optical pulses generated by Alice need to be matched with correct time tags. Here, we use the 500KHz green laser pulses to define a relative time reference. At Bob end, most of the green pulses will be detected by our APD (D5), of which the time is also recorded by the TDC and denoted as $t_{sync}$. Between any two adjacent synchronization pulses, there are 100 signal pulses transmitted by Alice (tagged as $s_0$, $s_1$, $s_2$, … $s_{99}$). Due to the time jitter of the APD (250ps) and the limited time resolution of the TDC (64ps), the detected events by Bob may deviate slightly from the base time and mix with some noise as well. Thus, we set a time window of $T_{window}$=5ns, and ascertain the time tag of any recorded event $t_{sig}$ by the following algorithm: 
\begin{equation}
\begin{split}
&n=\lfloor \frac{t_{sig}-t_{sync}+\triangle t}{T} \rfloor\\
\text{if, }&mod(\frac{t_{sig}-t_{sync}+\triangle t}{T})\leq T_{window}
\end{split}
\end{equation}
wherein $t_{sync}$ represents the adjacent synchronization event before $t_{sig}$ and $T=20ns$ is the signal period. The pre-offset of the signal time is set to be $\triangle t=2.5ns$. If the event is within the time window, it will be regarded as a valid signal bit with sequence number $n$, otherwise it will be discarded as noise. In this way we obtain the relative time tag $s’_n$ (see Extended Fig.2).

Owing to the high loss of the underwater channel, the sync pulses become single photon level when they arrive at the receiving end. Thus, we use the on-off type APD as detector for its high sensibility, which also means some of the pulses can be missed. To solve this problem, we designed a self-adapting algorithm, utilizing the accurate periodicity of the sync pulses to automatically recover the missed pulses as well as filtering out the error detections caused by random noise (see Extended Fig.3). In our experiment, only about 200K of the signals can be detected, by which we perfectly recover the origin 500KHz sync signals. Combining with relative time reference mentioned before, the time synchronization for the whole system is completed.

\section{Acknowledgments}
The authors thank Bill Munro and Jian-Wei Pan for helpful discussions. This research was supported by the National Key R\&D Program of China (2019YFA0308700, 2017YFA0303700), the National Natural Science Foundation of China (61734005, 11761141014, 11690033), the Science and Technology Commission of Shanghai Municipality (STCSM) (17JC1400403), and the Shanghai Municipal Education Commission (SMEC) (2017-01-07-00-02- E00049). X.-M.J. acknowledges additional support from a Shanghai talent program.

\bibliographystyle{apsrev4-1}
%

\renewcommand \thefigure{\arabic{figure}}
\setcounter{figure}{0}
\begin{figure*}[!]
\includegraphics[width=2\columnwidth]{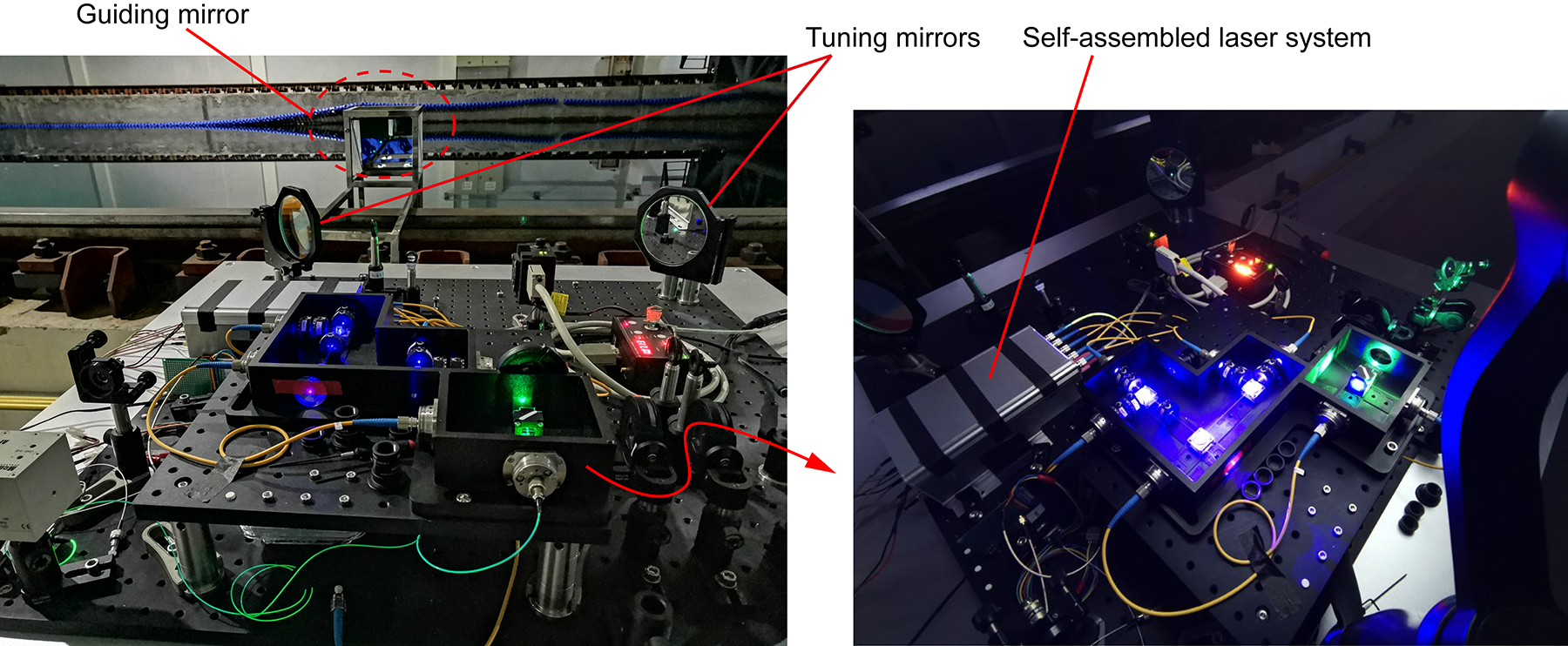}
\renewcommand{\figurename}{Extended Fig.}
\caption{\textbf{The QKD transmitter.}}
\label{extended_fig1}
\end{figure*}
\begin{figure*}[!]
\includegraphics[width=2\columnwidth]{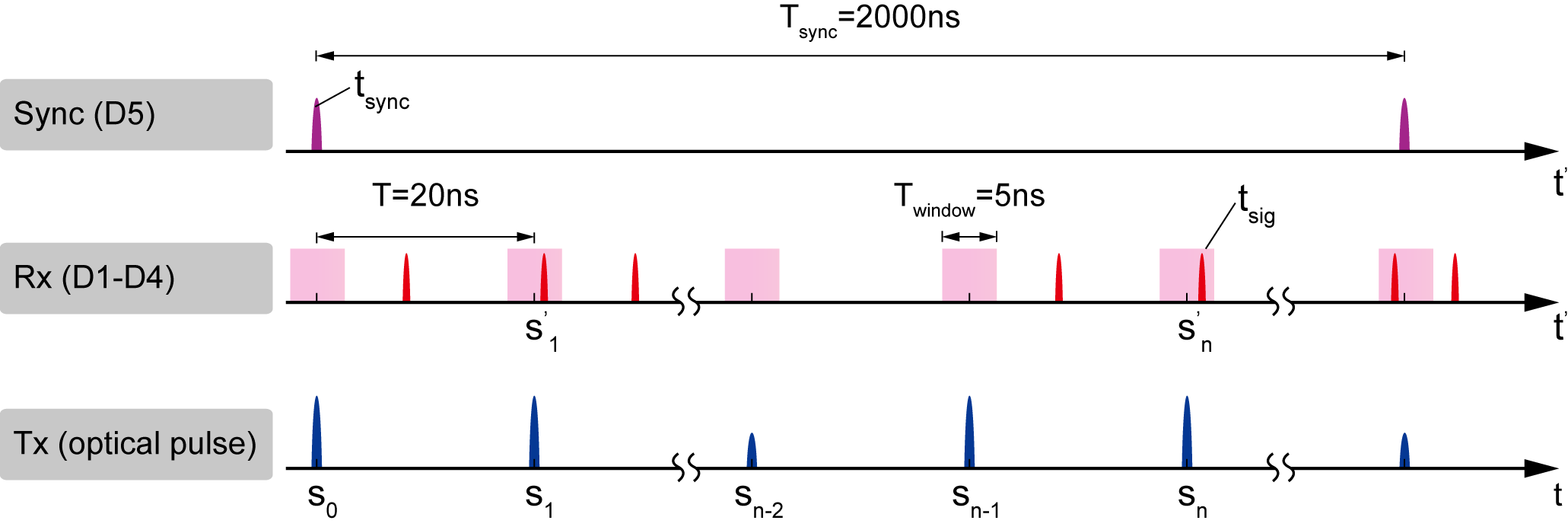}
\renewcommand{\figurename}{Extended Fig.}
\caption{\textbf{Bit sequence.}}
\label{extended_fig2}
\end{figure*}
\begin{figure*}[!]
\includegraphics[width=2\columnwidth]{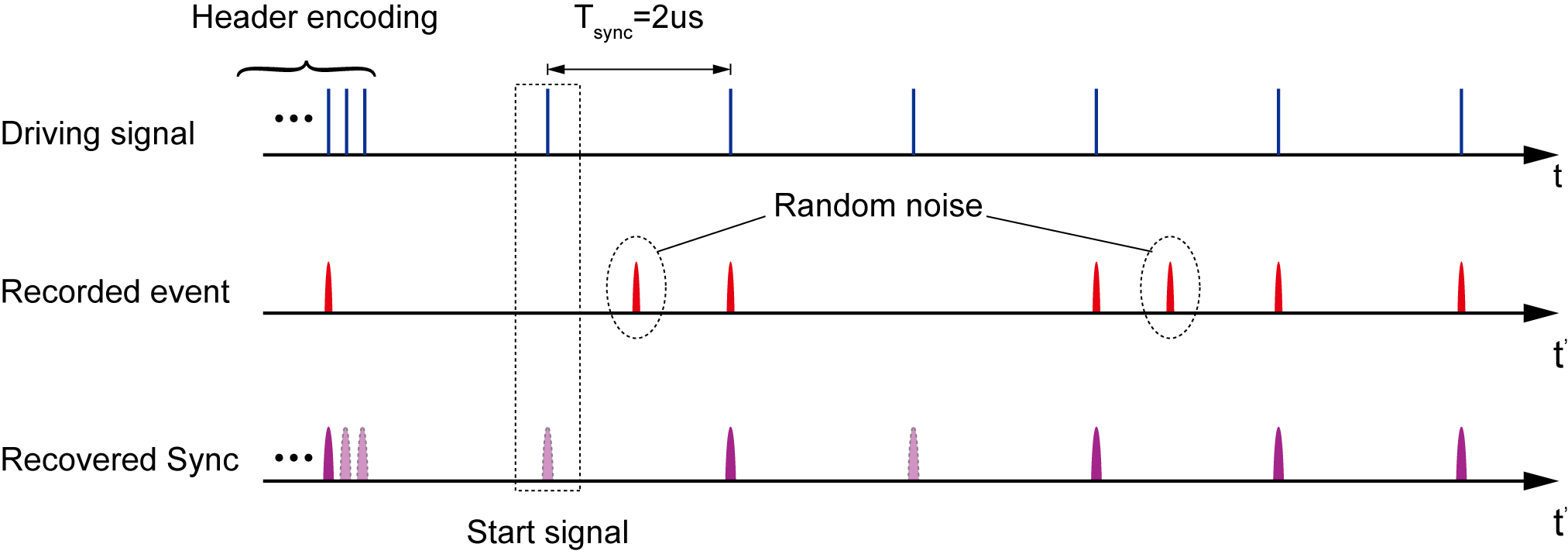}
\renewcommand{\figurename}{Extended Fig.}
\caption{\textbf{Sync sequence.}}
\label{extended_fig3}
\end{figure*}

\end{document}